\documentclass[%
 reprint,
 superscriptaddress,
 amsmath,amssymb,
 aps, physrev,
]{revtex4-2}

\usepackage{graphicx}
\usepackage{dcolumn}
\usepackage{bm}
\usepackage[unicode]{hyperref}

\usepackage{booktabs}
\usepackage[table]{xcolor}
\newcommand{\auditmark}{\textcolor{teal!70!black}{\ensuremath{\blacksquare}}}
\newcommand{\auditblank}{\ensuremath{\phantom{\blacksquare}}}
\colorlet{notcontactedbg}{teal!10}

\usepackage[round-mode=figures,round-precision=2,separate-uncertainty=true]{siunitx}

\usepackage[most]{tcolorbox}

\usepackage{multirow}
\usepackage{cancel}
\makeatletter
\newcommand{\balanceandclearpage}{%
  \par
  \close@column@grid
  \clearpage
  \twocolumngrid
}
\makeatother

 \let\epsilon\varepsilon
 \let\phi\varphi
 \let\theta\vartheta
\begin{document}

\preprint{APS/123-QED}

\title{\textbf{When Matching Observables Conceal Different Nonlinear Dynamics}}

\author{Tim De Coster}%
\email{tim.de-coster@ds.mpg.de}
\affiliation{Max Planck Institute for Dynamics and Self-Organisation, G\"ottingen, Germany}

\date{\today}

\begin{abstract}
Computational models are increasingly central to biomedical research, yet verifying their implementation often relies on a limited set of observables. I demonstrate that such tests can fail to detect inconsistencies with substantial consequences for nonlinear spatiotemporal dynamics. Using the three-variable Fenton-Karma model of cardiac excitation, I identify a difference in parameter conventions between two mathematically equivalent formulations. Combining equations and parameter values from different formulations without the required transformation produces an unintended dynamical system. Across eight published parameter sets, I find that conventional single-cell stimulation protocols can fail to reveal the mismatch, with action-potential-duration and conduction-velocity restitution remaining nearly identical. Nevertheless, the mismatch substantially alters spiral-wave core size, meandering, transient-chaos lifetimes, and the tissue-size--diffusion regimes supporting prolonged activity. These findings demonstrate that agreement in conventional observables does not establish dynamical equivalence and highlight the need for verification protocols tailored to the nonlinear dynamics of interest.
\end{abstract}

\maketitle


\section{\label{sec:introduction}Introduction}

As biomedical research increasingly relies on computational approaches, ensuring the reproducibility and reliability of numerical results has become an important challenge~\cite{zhong2026reproducibility}. Reproducibility requires not only transparent reporting of model equations and parameters but also reliable procedures for assessing their numerical implementation. Such assessments commonly rely on comparisons with established results using a limited set of observables. However, agreement in these observables does not necessarily establish equivalence of the underlying dynamics.

Cardiac electrophysiology provides a particularly relevant example. Reduced reaction--diffusion models, such as the three-variable Fenton-Karma model, are widely used to investigate cardiac excitation, wave propagation, and spiral-wave dynamics~\cite{fenton1998vortex,fenton2002multiple}. Their implementations are frequently assessed using single-cell action potentials, restitution curves, and one-dimensional conduction properties. Although these observables characterize important aspects of excitability, they probe only a restricted subset of the possible dynamics. Discrepancies may therefore remain undetected under standard protocols while becoming apparent under different stimulation conditions or in more complex spatial configurations. Moreover, mathematically equivalent model formulations may employ different parameter conventions, making their consistent interpretation essential for reproducible simulations.

Here, I demonstrate that mixing parameter conventions from two published formulations of the Fenton-Karma model produces a distinct dynamical system whose differences can remain inconspicuous in conventional observables while substantially altering spiral-wave dynamics and the persistence of transient spatiotemporal chaos. A survey of recent publications further reveals incompatible or insufficiently specified parameter conventions. Together, these findings expose a limitation of conventional verification procedures and highlight the need for unambiguous model specifications and tests tailored to the dynamics of interest.

\section{Methods and results}
\subsection{General principle}

Consider a general class of nonlinear dynamical systems described by

\begin{equation}
\dot{\mathbf{u}}=\mathbf{F}(\mathbf{u};\boldsymbol{\theta}),
\end{equation}

where \(\mathbf{u}\in\mathbb{R}^n\) denotes the state vector and \(\boldsymbol{\theta}\) represents the model parameters. We consider two formulations of the same model, denoted by \(A\) and \(B\), which differ in their parameter conventions. Their governing equations are given by

\begin{equation}
\mathbf{F}_i(\mathbf{u};\boldsymbol{\theta}),
\qquad i\in\{A,B\}.
\end{equation}

The two formulations are mathematically equivalent if there exists a parameter transformation \(\mathbf{T}\) such that

\begin{equation}
\mathbf{F}_A(\mathbf{u};\boldsymbol{\theta}) = \mathbf{F}_B(\mathbf{u};\mathbf{T}(\boldsymbol{\theta}))
\end{equation}

for all admissible states and parameter values. Consequently, applying identical numerical parameter values to both formulations without accounting for this transformation may result in different dynamical systems, even though the formulations describe the same underlying model.

The corresponding spatially extended systems are governed by

\begin{equation}
\frac{\partial\mathbf{u}}{\partial t}
=
\mathbf{F}_i(\mathbf{u};\boldsymbol{\theta})
+
\mathbf{D}\nabla^2\mathbf{u},
\qquad i\in\{A,B\},
\end{equation}

where \(\mathbf{D}\) is a constant diffusion matrix, assumed to be identical for both formulations.

To assess the agreement between the two implementations, we introduce a dimension-dependent observation operator \(\mathcal{O}_d\), where \(d=0\) denotes the spatially homogeneous system and \(d=1,2\) its one- and two-dimensional spatial extensions, respectively. Let \(\mathbf{u}_i^{(d)}\) denote the solution obtained from formulation (i) in dimension (d), using identical numerical parameter values and otherwise equivalent simulation conditions. The discrepancy between the resulting observables is defined as

\begin{equation}
\Delta_d
=
\left|
\mathcal{O}_d\left[\mathbf{u}_A^{(d)}\right]
-
\mathcal{O}_d\left[\mathbf{u}_B^{(d)}\right]
\right|_d,
\qquad d\in{0,1,2},
\end{equation}

where \(|\cdot|\) denotes an appropriate norm on the space of observables. The discrepancy \(\Delta_d\) depends on the chosen observation operator and need not characterize differences in the full dynamics. Of particular interest is the scenario in which

\begin{equation}
\boxed{
\Delta_0\approx 0,\qquad
\Delta_1\approx 0,\qquad
\Delta_2\gg 0
}
\label{eq:observable_discrepancy}
\end{equation}

where discrepancies are understood relative to the characteristic scales of the respective observables. In this scenario, agreement in selected lower-dimensional observables conceals substantial differences in higher-dimensional dynamics.

\begin{figure*}
\includegraphics[width=\textwidth]{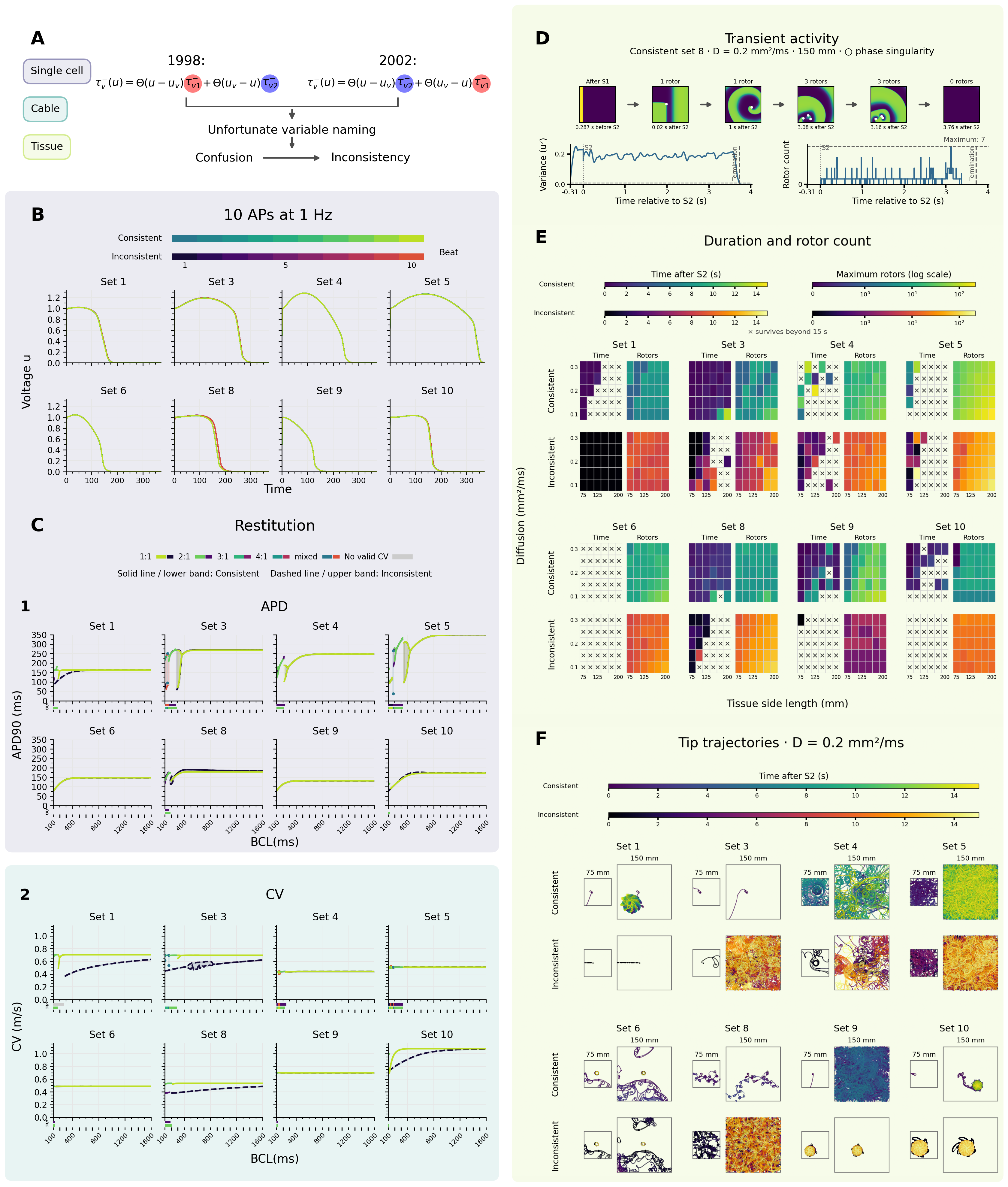}
\caption{\label{fig:overview}
        \textbf{Fenton-Karma simulations showing that matching observables can conceal different dynamics.} (\textbf{A}) Parameter-convention mapping between the two Fenton-Karma formulations. (\textbf{B}) Single-cell action potentials. (\textbf{C}) APD and CV restitution. (\textbf{D}) Representative transient spiral-wave activity and phase-singularity evolution. (\textbf{E}) Termination-time and phase-singularity statistics over tissue size and diffusion. (\textbf{F}) Spiral-tip trajectories. Consistent (blue-yellow) and inconsistent (black-red) implementations can be distinguished by their respective colour schemes in every panel.
        }
\end{figure*}

\subsection{Concrete example}

\subsubsection{Model and parameter conventions}

Let's consider the three-variable Fenton-Karma model of cardiac excitation~\cite{fenton1998vortex,fenton2002multiple}. The dimensionless transmembrane voltage \(u\) evolves according to
\begin{equation}
    \frac{\partial u}{\partial t}
    =
    D\nabla^2u
    -
    \left(
        J_{\mathrm{fi}}+
        J_{\mathrm{so}}+
        J_{\mathrm{si}}
    \right),
    \label{eq:fk}
\end{equation}
coupled to evolution equations for the recovery variables \(v\) and \(w\). Here, \(D\) is the diffusion coefficient, and \(J_{\mathrm{fi}}\), \(J_{\mathrm{so}}\), and \(J_{\mathrm{si}}\) denote the fast inward, slow outward, and slow inward currents, respectively. I use the standard Fenton-Karma normalization, with a membrane capacitance of \(C_m=1\,\mu\mathrm{F}/\mathrm{cm}^2\).

The original 1998 publication~\cite{fenton1998vortex} introduced three parameter sets, while the subsequent 2002 publication~\cite{fenton2002multiple} examined a broader range of dynamical regimes and reported ten parameter sets. Although both publications use the same parameter names, their formulations differ in the assignment of the recovery time constants \(\tau_{v1}^{-}\) and \(\tau_{v2}^{-}\) to the corresponding voltage regimes.

In the 1998 formulation, the voltage-dependent recovery time scale is defined as
\begin{equation}
    \tau_{v,1998}^{-}(u)
    =
    \Theta(u-u_v)\tau_{v1}^{-}
    +
    \Theta(u_v-u)\tau_{v2}^{-},
    \label{eq:tau1998}
\end{equation}
where \(\Theta\) is the Heaviside function and \(u_v\) denotes the switching threshold. The 2002 formulation instead uses
\begin{equation}
    \tau_{v,2002}^{-}(u)
    =
    (1-q)\tau_{v1}^{-}
    +
    q\tau_{v2}^{-},
    \qquad
    q=
    \begin{cases}
        1, & u\geq u_v,\\
        0, & u<u_v.
    \end{cases}
    \label{eq:tau2002}
\end{equation}
Fig.~\ref{fig:overview}A presents both definitions in a common notation, making the interchange of the two time constants explicit.

The formulations are mathematically equivalent under a permutation of these parameters. Defining
\begin{equation}
    \boldsymbol{\tau}^{-}
    =
    \begin{bmatrix}
        \tau_{v1}^{-}\\
        \tau_{v2}^{-}
    \end{bmatrix},
    \qquad
    \mathbf{A}
    =
    \begin{bmatrix}
        0 & 1\\
        1 & 0
    \end{bmatrix},
\end{equation}
the transformation \(\mathbf{T}\) acts as \(\boldsymbol{\tau}^{-}\mapsto\mathbf{A}\boldsymbol{\tau}^{-}\), leaving all other parameters unchanged. The corresponding reaction terms therefore satisfy
\begin{equation}
    \mathbf{F}_{1998}
    (\mathbf{u};\boldsymbol{\theta})
    =
    \mathbf{F}_{2002}
    (\mathbf{u};\mathbf{T}(\boldsymbol{\theta})).
    \label{eq:fk_equivalence}
\end{equation}

This distinction becomes relevant when parameter sets from the 2002 publication are used in an implementation based on the 1998 equations. Although the parameter names are identical, the two recovery time constants must be exchanged to preserve the intended dynamics (Fig.~\ref{fig:overview}A). Without this transformation,
\begin{equation}
    \mathbf{F}_{1998}
    (\mathbf{u};\boldsymbol{\theta})
    \neq
    \mathbf{F}_{2002}
    (\mathbf{u};\boldsymbol{\theta})
    \label{eq:fk_mismatch}
\end{equation}
in general. Thus, directly transferring the published parameter values between the two formulations can produce a different dynamical system.

\subsubsection{Observables and numerical methods}

I investigated this inconsistency using eight of the ten parameter sets reported in the 2002 publication~\cite{fenton2002multiple} (sets 2 and 7 were skipped because there \(\tau_{v1}^{-} = \tau_{v2}^{-}\)). For each set, I compared the intended dynamics, obtained with consistent parameter conventions (named \textit{consistent}), against those resulting from direct assignment of the same numerical values to the 1998 formulation without exchanging \(\tau_{v1}^{-}\) and \(\tau_{v2}^{-}\) (named \textit{inconsistent}). All other parameters and simulation conditions were identical.

I evaluated single-cell action potentials (10 pulses at \(1\,\mathrm{Hz}\)), action-potential-duration (APD) restitution, and conduction-velocity (CV) restitution as conventional electrophysiological observables. I then examined two-dimensional wave dynamics, including spiral-wave activity and transient spatiotemporal chaos, across different domain sizes and diffusion coefficients. Spiral-wave dynamics were characterized using phase-singularity trajectories and statistics.

Simulations were performed using the \texttt{pigreads} Python package~\cite{kabus2026pigreads}, with a fixed time step of \(\Delta t=0.05\,\mathrm{ms}\) and spatial grid spacing of \(\Delta x=0.25\,\mathrm{mm}\). Unless otherwise specified, the diffusion coefficient was \(D=0.25\,\mathrm{mm}^2/\mathrm{ms}\); additional values were considered in parameter sweeps.

S1 and S2 stimuli were implemented as instantaneous resets of the dimensionless voltage to \(u=0.5\), ensuring excitation across the investigated parameter sets. In two-dimensional simulations, the S2 stimulus was applied to one quarter of the domain when the voltage at its centre fell below \(u=0.5\) following passage of the initial wave. Two-dimensional simulations were performed over a \(15\,\mathrm{s}\) observation window, with activity considered terminated when the spatial voltage variance fell below \(0.01\).

Phase singularities were detected using a hybrid phase-winding and isoline-intersection method~\cite{iyer2001experimentalist,jacquemet2018phase}. The local phase was defined as
\begin{equation}
\phi=\operatorname{atan2}\left(\frac{v-v^\ast}{0.6},u-0.5\right),
\end{equation}
where \(v^\ast\) was chosen separately for each parameter set from within the action-potential loop at \(u=0.5\), minimizing the maximum wrapped phase difference between neighboring grid points in representative simulations. Candidate singularities were identified by phase winding of \(\pm2\pi\) around individual grid cells and localized with subgrid accuracy by bilinear interpolation of the intersecting isolines \(u=0.5\) and \(v=v^\ast\). When multiple intersections occurred, the root consistent with the winding charge and having the largest Jacobian magnitude was retained. The resulting positions were used to reconstruct spiral-tip trajectories.

\subsection{Observable agreement and dynamical discrepancies}

Despite their different recovery kinetics, the consistent and inconsistent implementations produced nearly indistinguishable single-cell action potentials under the stimulation protocols examined [Fig.~\ref{fig:overview}B], corresponding to small \(\Delta_0\) for this observable. Quantitative differences were present but small, with discrete \(L_2\) differences of approximately \(1\) or \(10^{-5}\) (see Fig.~\ref{fig:APdiff}A in End Matter \ref{app:validation}). Restitution measurements were more sensitive: the discrepancies in APD (\(d=0\)) and CV (\(d=1\)) restitution were larger for several parameter sets, although these differences were sometimes difficult to discern visually [Fig.~\ref{fig:overview}C; see also Fig.~\ref{fig:APdiff}B in End Matter \ref{app:validation}]. For sets 4, 5, 6, and 9, the restitution curves remained particularly similar, demonstrating that conventional excitation and propagation measurements do not consistently reveal the parameter mismatch.

The consequences were considerably more pronounced in two-dimensional tissue, where spiral-wave observables exhibited substantial discrepancies (\(\Delta_2\)). Spiral-wave core size, meandering trajectories, phase-singularity dynamics, and activity termination times depended strongly on the parameter convention [Fig.~\ref{fig:overview}D-F; see also Fig.~\ref{fig:tiptraces} and Fig.~\ref{fig:rotorframes} in End Matter \ref{app:tiptracing}]. Parameter set 9 provides a particularly clear realization of Eq.~\eqref{eq:observable_discrepancy}: despite negligible discrepancies in APD and CV restitution (\(\Delta_0,\Delta_1\)), the mean observed activity duration across the investigated tissue sizes and diffusion coefficients increased from \(4.19\) to \(14.51\,\mathrm{s}\) under the switched convention (\(\Delta_2\)). More generally, the switched implementation supported activity throughout the \(15\,\mathrm{s}\) observation window over a larger portion of the investigated tissue-size--diffusion plane for several parameter sets. This trend was not universal: for set 1, the mean observed activity duration instead decreased from \(10.14\) to \(0.12\,\mathrm{s}\). Set 6 exhibited activity throughout the \(15\,\mathrm{s}\) observation window under both conventions for all investigated configurations, while showing comparatively minor differences in spiral-wave dynamics. Thus, agreement in local excitation, restitution, or even observed activity duration did not necessarily imply agreement in the underlying spatiotemporal dynamics.

To assess the prevalence of this issue, I systematically reviewed 30 eligible publications using the relevant Fenton-Karma formulations over the preceding five years. Ten (33.3\%) reported incompatible equation-parameter conventions, five (16.7\%) reported consistent conventions, and fifteen (50.0\%) provided insufficient information to determine the convention used. Thus, among the 15 publications with classifiable descriptions, two-thirds (66.7\%) reported an inconsistency. Examination of available source code identified five instances of the mismatch and six consistent implementations despite insufficient methodological reporting. The authors of 12 publications were contacted, with responses pending at the time of assessment. The complete classification and code audit are provided in Table~\ref{tab:audit} of End Matter \ref{app:litaudit}.

\section{Discussion}

These findings reveal a blind spot in conventional verification of nonlinear reaction--diffusion models: small discrepancies in selected lower-dimensional observables (\(\Delta_0,\Delta_1\)) can coexist with substantial discrepancies in spatial dynamics (\(\Delta_2\)). That this limitation already manifests in a three-variable model is particularly noteworthy, given the additional degrees of freedom and interacting processes in more complex models. This is particularly relevant for applications involving reentrant activity, where spiral-wave trajectories, breakup, and termination may influence model-based investigations of arrhythmia mechanisms or strategies for terminating reentry, including the evaluation of ablation lesions. Verification protocols should therefore include observables relevant to the intended application. Importantly, the mismatch is not intrinsically undetectable in lower-dimensional simulations; alternative stimulation protocols may also reveal it.

The literature survey identifies two related obstacles to reproducibility: incompatible equation--parameter specifications and insufficient methodological reporting. An inconsistency in a published description does not establish that the corresponding simulation code was mismatched, while an implementation may be consistent despite incomplete reporting. These findings do not, by themselves, invalidate previously reported scientific conclusions, but demonstrate the need for unambiguous model specifications. A minimal reporting requirement would be to identify the equation source, parameter-set source, and any transformation applied between them. For example: ``We used the Fenton-Karma equations of Ref.~\cite{fenton1998vortex} with parameter set 8 of Ref.~\cite{fenton2002multiple}, exchanging \(\tau_{v1}^{-}\) and \(\tau_{v2}^{-}\) to reconcile the parameter conventions.''

More broadly, agreement in selected observables does not establish dynamical equivalence. Reliable computational studies require explicit equation--parameter correspondence and verification against observables that probe the phenomena under investigation. The present example illustrates how an apparently minor implementation inconsistency can remain inconspicuous under conventional checks while substantially altering nonlinear spatial dynamics.

\begin{acknowledgments}
\noindent I would like to thank Ulrich Parlitz, Peer Drescher and Benjamin Weiss for inspiring discussions. During the preparation of this work I used ChatGPT in order to help with writing the manuscript. After using this tool, I reviewed and edited as needed and take full responsibility for the content of this article.\\

\noindent This work was supported by the European Union's Horizon Europe research and innovation programme (under the Marie Skłodowska-Curie grant agreement No 101205806 to TDC).
\end{acknowledgments}

\section*{Data Availability}
\noindent The data that support the findings of this article are openly available \cite{CodeFKConfusion}.

\bibliographystyle{apsrev4-1}
\bibliography{main}

\balanceandclearpage

\section*{End Matter}\label{appendix}

\appendix
\makeatletter
\renewcommand{\appendixname}{}
\makeatother

\section{Classical validation test results}\label{app:validation}
\begin{figure}[b]
\includegraphics[width=\columnwidth]{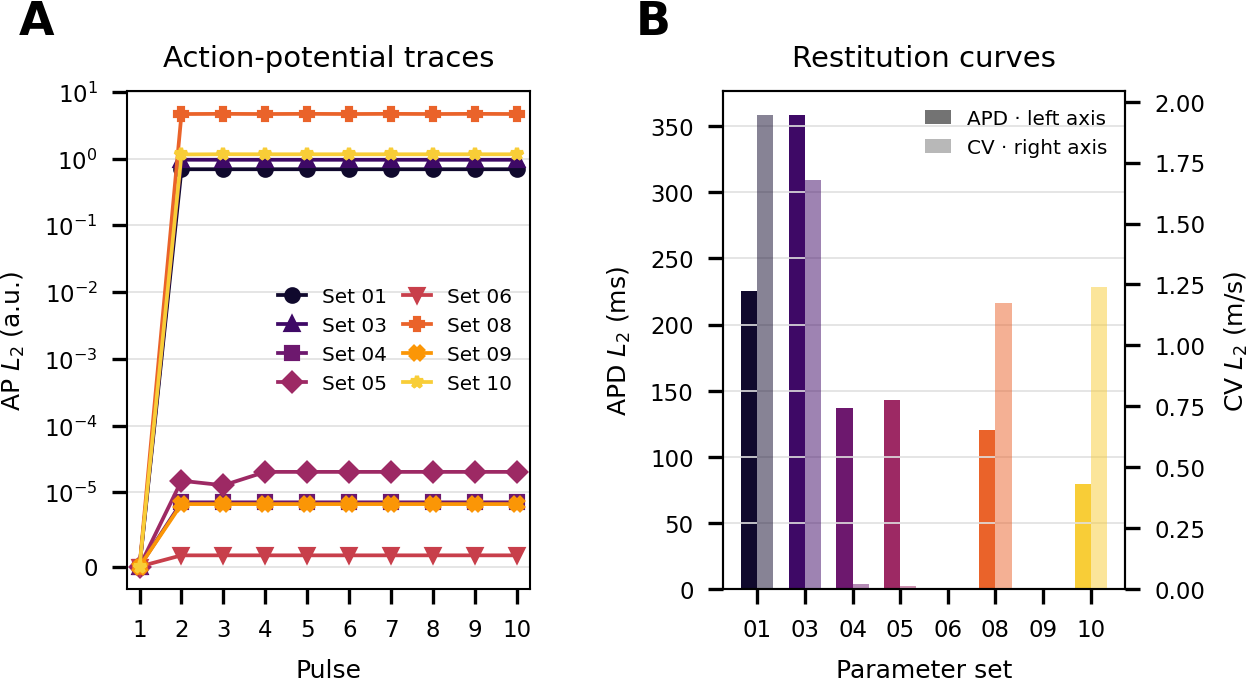}
\caption{\label{fig:APdiff}
        \textbf{Quantification of the difference between parameter sets APs and restitution curves.} (\textbf{A}) AP differences over 10 consecutive pulses for each couple of parameter sets. (\textbf{B}) APD and CV restitution differences for the same couple of parameter sets.
        }
\end{figure}

APs and restitution curves, the classical validation metrics, usually are only looked at by comparing with a figure from another journal, and not by numerical details because they are usually not available (don't worry, I made mine available under \texttt{data/cells} in \cite{CodeFKConfusion} such that from now on, this can be used as a hard numerical reference). This also allows for numerical quantification of the differences between all measurements, which are visualised in Fig.~\ref{fig:APdiff}. For both the AP-difference and the restitution curve difference, I made use of the \(L_2\) norm defined as
\begin{align}
    AP\,L_{2} &= \sqrt{\sum_{t=0ms}^{1000 ms}{(AP_{consist.}-AP_{inconsist.})^{2}}},\\
    X\,L_{2} &= \sqrt{\sum_{BCL=1600ms}^{100 ms}{(X_{consist.}-X_{inconsist.})^{2}}},
\end{align}
for \(X=APD/CV\). Note that AP differences stabilize quickly after beat 1 and are always barely noticeable, while restitution curve differences can be very high. However, their visual inspection in Fig.~\ref{fig:overview}C1-2 might not indicate that. This is due to the chaotic regimes at low cycle lengths which can amplify the \(L_{2}\) norm quickly.

\section{Phase-singularity dynamics}\label{app:tiptracing}

\begin{figure}[b]
\includegraphics[width=\columnwidth]{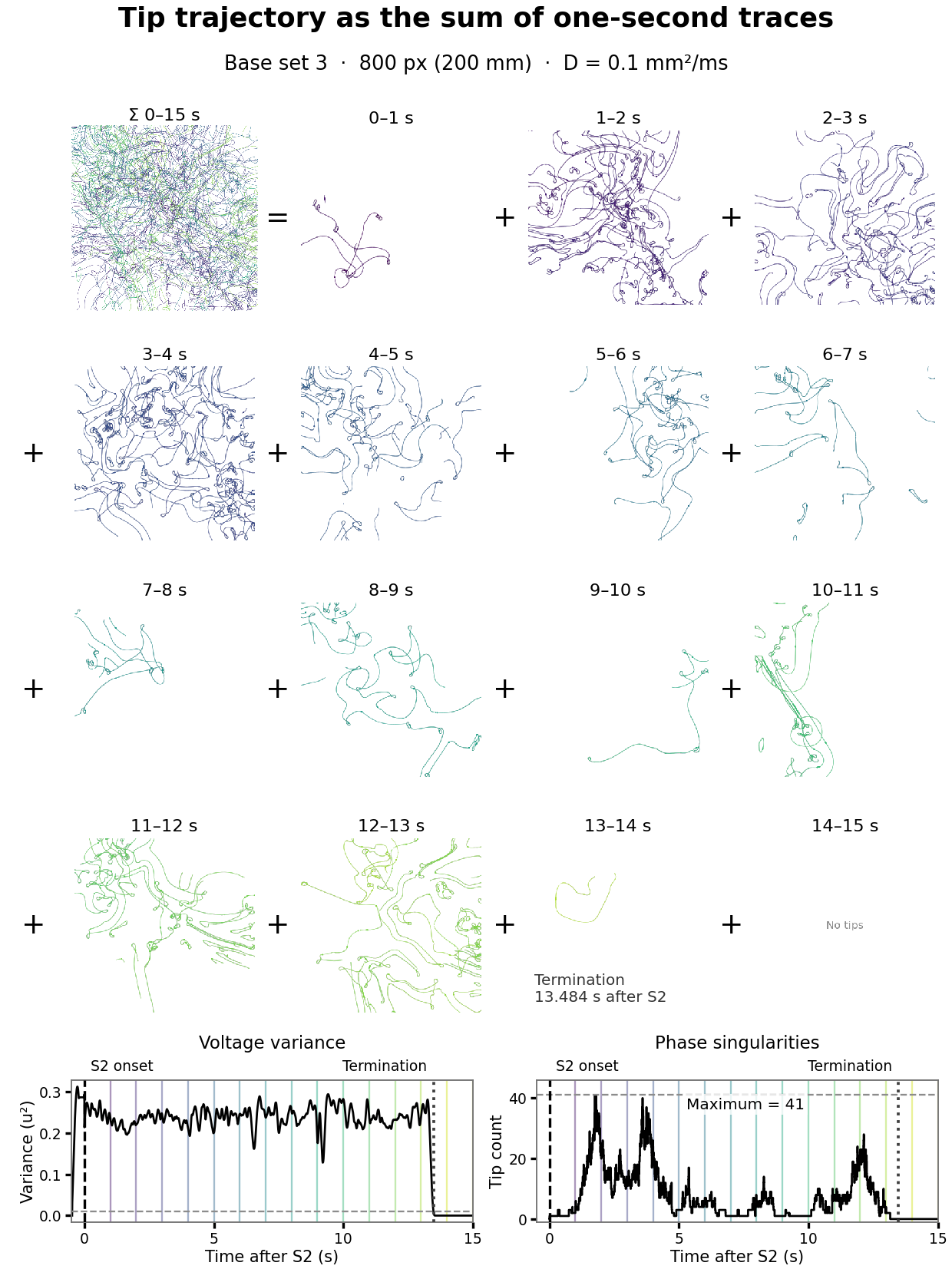}
\caption{\label{fig:tiptraces}
    \textbf{Phase-singularity dynamics preceding termination of spatiotemporal chaos.}
    Trajectories are shown in \(1\,\mathrm{s}\) intervals, with color indicating temporal evolution. Spatial complexity increases around \(t=4\,\mathrm{s}\), decreases around \(t=10\,\mathrm{s}\), and increases again around \(t=13\,\mathrm{s}\) before termination during the 14th second.
}
\end{figure}

Phase-singularity positions, spatial voltage variance, and pulse counts were recorded throughout each simulation. Visual inspection of selected simulations revealed an increase in phase-singularity count shortly before termination, indicating a transient increase in spatial complexity preceding collapse. Fig.~\ref{fig:tiptraces} illustrates this evolution in \(1\)s intervals, while Fig.~\ref{fig:overview} displays the superimposed trajectories over the entire observation window.

\begin{figure*}
\includegraphics[width=\textwidth]{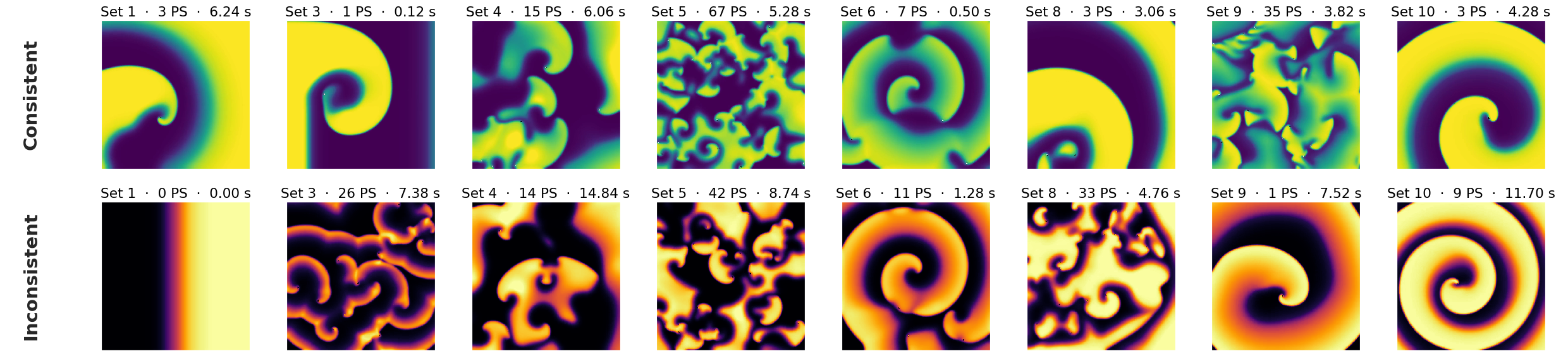}
\caption{\label{fig:rotorframes}
    \textbf{Representative voltage snapshots near maximum rotor count.}
    Consistent (top) and inconsistent (bottom) implementations are shown for all investigated parameter sets at a tissue width of \(150\,\mathrm{mm}\) and \(D=0.2\,\mathrm{mm}^2/\mathrm{ms}\). The planar wavefront for the inconsistent implementation of set 1 indicates unsuccessful spiral-wave initiation. Rotor counts and snapshot times are indicated above each panel.
}
\end{figure*}

While phase-singularity trajectories characterize the temporal evolution and spatial extent of spiral-wave activity, voltage snapshots provide complementary information on wavefront morphology and spatial organization. Representative voltage snapshots near the maximum rotor count are shown in Fig.~\ref{fig:rotorframes}, allowing direct visual comparison between consistent and inconsistent implementations under identical spatial conditions.

The complete tip trajectories, associated statistics, plus simulation videos for all 480 configurations (2 variants \(\times\) 8 parameter sets \(\times\) 6 sizes \(\times\) 5 diffusion coefficients) investigated in Fig.~\ref{fig:overview}E are publicly available as a dataset for future studies~\cite{CodeFKConfusion}.

\begin{table*}
\caption{Audit of Fenton-Karma model specifications in 30 recent publications. Filled squares indicate the applicable categories. Published descriptions are classified exclusively as consistent, inconsistent, or insufficiently specified (``Insuff.''). Code classifications are not mutually exclusive, as individual publications may employ multiple implementations. ``Unavail.'' indicates confirmed unavailability of the relevant code. The final column records the code source or contact status. The symbol \(\dagger\) denotes code obtained after the initial public release of this article's preprint.}
\label{tab:audit}
\centering
\setlength{\tabcolsep}{2.8pt}
\renewcommand{\arraystretch}{1.04}
\scriptsize
\begin{tabular}{@{}r l ccc
    !{\hspace{3pt}\vrule width 1.3pt\hspace{3pt}}
    ccc l@{}}
\toprule
 & & \multicolumn{3}{c}{Published description}
   & \multicolumn{3}{c}{Code inspection}
   & Code source / status \\
\cmidrule(lr){3-5}
\cmidrule(lr){6-8}
Ref. & Publication
     & Consist. & Inconsist. & Insuff.
     & Consist. & Inconsist. & Unavail. & \\
\midrule
\cite{fenton1998vortex} & [1998, Fenton \& Karma]
  & \auditmark & \auditblank & \auditblank
  & \multicolumn{4}{>{\columncolor{notcontactedbg}}c}{} \\

\cite{fenton2002multiple} & [2002, Fenton et al.]
  & \auditmark & \auditblank & \auditblank
  & \multicolumn{4}{>{\columncolor{notcontactedbg}}c}{%
      \multirow{-2}{*}{%
          \textcolor{teal!60!black}{\textit{Original papers}}%
      }%
    } \\
\midrule

\cite{li2025sonogenetics} & [2025, Li et al.]
  & \auditmark & \auditblank & \auditblank
  & \multicolumn{4}{>{\columncolor{notcontactedbg}}c}{} \\
\cite{von2025analysing} & [2025, Von Koeller et al.]
  & \auditmark & \auditblank & \auditblank
  & \multicolumn{4}{>{\columncolor{notcontactedbg}}c}{} \\
\cite{luo2024unpinning} & [2024, Luo et al.]
  & \auditmark & \auditblank & \auditblank
  & \multicolumn{4}{>{\columncolor{notcontactedbg}}c}{} \\
\cite{shi2023information} & [2023, Shi et al.]
  & \auditmark & \auditblank & \auditblank
  & \multicolumn{4}{>{\columncolor{notcontactedbg}}c}{} \\
\cite{dantasenc} & [2022, Dantas et al.]
  & \auditmark & \auditblank & \auditblank
  & \multicolumn{4}{>{\columncolor{notcontactedbg}}c}{%
      \multirow{-5}{*}{%
          \textcolor{teal!60!black}{\textit{Not contacted}}%
      }%
    } \\

\cite{echeverria2025boundary} & [2025, Echevarria-Alar \& Rappel]
  & \auditblank & \auditmark & \auditblank
  & WebGL & MATLAB & \auditblank 
  & Public repository \\
\cite{mulimani2025spiral} & [2025, Mulimani \& Rappel]
  & \auditblank & \auditmark & \auditblank
  & \auditblank & \auditmark & \auditblank
  & Author-provided \\
\cite{dix2024influence} & [2024, Dix et al.]
  & \auditblank & \auditmark & \auditblank
  & \auditblank & \auditmark & \auditblank
  & Author-provided \\
\cite{aron2023optimising} & [2023, Aron et al.]
  & \auditblank & \auditmark & \auditblank
  & \auditblank & \auditblank & \auditblank
  & Pending response \\
\cite{hu2023polarized} & [2023, Hu et al.]
  & \auditblank & \auditmark & \auditblank
  & \auditblank & \auditmark & \auditblank
  & Author-provided \\
\cite{li2023reordering} & [2023, Li et al.]
  & \auditblank & \auditmark & \auditblank
  & Fig. 11 (3D) & Fig. 10 (2D) & \auditblank
  & Author-provided \\
\cite{nieto2023bayesian} & [2022, Nieto et al.]
  & \auditblank & \auditmark & \auditblank
  & \auditblank & \auditblank & \auditblank
  & Pending response \\
\cite{rappel2022intermittent} & [2022, Rappel]
  & \auditblank & \auditmark & \auditblank
  & \auditblank & \auditblank & \auditblank
  & Pending response \\
\cite{steyer2023role} & [2022, Steyer et al.]
  & \auditblank & \auditmark & \auditblank
  & \auditblank & \auditblank & \auditmark
  & Author response \\
\cite{marcotte2021robust} & [2021, Marcotte et al.]
  & \auditblank & \auditmark & \auditblank
  & \auditmark & \auditblank & \auditblank
  & Author-provided \\

\cite{bury2026reinforcement} & [2026, Bury et al.]
  & \auditblank & \auditblank & \auditmark
  & \auditmark & \auditblank & \auditblank
  & Author-provided \\
\cite{angelaki2024t} & [2024, Angelaki et al.]
  & \auditblank & \auditblank & \auditmark
  & \auditblank & \auditblank & \auditblank
  & Pending response \\
\cite{delshad2025predicting} & [2025, Delshad et al.]
  & \auditblank & \auditblank & \auditmark
  & \auditblank & \auditblank & \auditblank
  & Pending response \\
\cite{kaboudian2024fast} & [2024, Kaboudian et al.]
  & \auditblank & \auditblank & \auditmark
  & \auditblank & \auditblank & \auditblank
  & Pending response \\
\cite{ntagiantas2024estimation} & [2024, Ntagiantas et al.]
  & \auditblank & \auditblank & \auditmark
  & \auditblank & \auditblank & \auditblank
  & Pending response \\
\cite{ugarte2024quantifying} & [2024, Ugarte et al.]
  & \auditblank & \auditblank & \auditmark
  & \auditmark & \auditblank & \auditblank
  & Author-provided \\
\cite{ugarte2024variable} & [2024, Ugarte et al.]
  & \auditblank & \auditblank & \auditmark
  & \auditmark & \auditblank & \auditblank
  & Author-provided \\
\cite{dantas2022thermal} & [2022, Dantas et al.]
  & \auditblank & \auditblank & \auditmark
  & \auditblank & \auditblank & \auditblank
  & Pending response \\
\cite{orozco2022nonlinear} & [2022, Orozco et al.]
  & \auditblank & \auditblank & \auditmark
  & \auditmark & \auditblank & \auditblank
  & Author-provided \\
\cite{rappel2022stochastic} & [2022, Rappel et al.]
  & \auditblank & \auditblank & \auditmark
  & \auditblank & \auditblank & \auditblank
  & Pending response \\
\cite{shahi2022machine} & [2022, Shahi et al.]
  & \auditblank & \auditblank & \auditmark
  & \auditblank & \auditblank & \auditblank
  & Pending response \\
\cite{ugarte2022fractional} & [2022, Ugarte et al.]
  & \auditblank & \auditblank & \auditmark
  & \auditmark & \auditblank & \auditblank
  & Author-provided \\
\cite{uzelac2022spiral} & [2022, Uzelac et al.]
  & \auditblank & \auditblank & \auditmark
  & \auditblank & \auditblank & \auditblank
  & Pending response \\
\cite{orozco2021local} & [2021, Orozco et al.]
  & \auditblank & \auditblank & \auditmark
  & \auditmark & \auditblank & \auditblank
  & Author-provided \\
\cite{shahi2021long} & [2021, Shahi et al.]
  & \auditblank & \auditblank & \auditmark
  & \auditblank & \auditblank & \auditblank
  & Pending response \\

\bottomrule
\end{tabular}
\end{table*}

\section{Literature audit}\label{app:litaudit}

I classified the equation-parameter conventions of 30 eligible publications into three mutually exclusive categories based solely on their published model descriptions (Table~\ref{tab:audit}). Where the published convention was inconsistent or insufficiently specified, I examined publicly available code or contacted the authors.

For most inspected publications with inconsistent equation-parameter specifications, the same inconsistency was found in the code. However, consistent implementations were identified despite inconsistent descriptions, suggesting that some inconsistencies in the published text may be attributable to typographical errors.

The primary code audit reflects the information available at the initial public release of this article's preprint. Further contributions from authors are welcome and will be incorporated into subsequent revisions. For transparency, code received after public disclosure of the parameter-convention mismatch is marked separately.

\end{document}